\documentclass[11pt,onecolumn]{IEEEtran}

\usepackage{amsthm}
\usepackage{subfigure}
\usepackage{slashbox}
\usepackage{graphicx}
\usepackage{amssymb,amstext,amsmath}
\usepackage{float}
\usepackage{bbm}
\usepackage{bm}
\usepackage{multirow}
\usepackage{makecell}
\usepackage{color}
\usepackage{caption}
\usepackage{verbatim}

\usepackage{tikz}
\usetikzlibrary{arrows}

\newcommand{\x}{{\bf x}}

\newcommand{\normalized}{\widetilde}

\newtheorem{theorem}{Theorem}
\newtheorem{corollary}[theorem]{Corollary}

\def\Item$#1${\item $\displaystyle#1$
\hfill\refstepcounter{equation}(\theequation)}

\newcommand\blfootnote[1]{%
  \begingroup
  \renewcommand\thefootnote{}\footnote{#1}%
  \addtocounter{footnote}{-1}%
  \endgroup
}

\begin{document}

\title{Efficient Multiple Importance Sampling Estimators}

\author{V\'ictor Elvira,
         Luca Martino,
        David Luengo,
        and~M\'onica F. Bugallo
}


\maketitle
\blfootnote{V. Elvira is with the Dept. of Signal Theory and Communications, Universidad Carlos III de Madrid,  Legan\'es (Spain). L. Martino is with the Dept. of Mathematics and Statistics, University of Helsinki,  Helsinki (Finland). D. Luengo is with the Dept. of Signal Theory and Communications, Universidad Polit\'ecnica de Madrid, Madrid (Spain). M. F. Bugallo is with the Dept. of Electrical and Computer Engineering, Stony Brook University, New York (USA).}
\blfootnote{This work has been supported by the Spanish government's projects TEC2012-38800-C03-01, TEC2012-38883-C02-01, TEC2012-38058-C03-01, TEC2013-41718-R and S2010/BMD-2422; by the BBVA Foundation's project MG-FIAR; by ERC grant 239784 and AoF grant 251170; by the NSF's Award CCF-0953316; and by the EU's Marie Curie ITN MLPM2012 (Ref. 316861).}

\begin{abstract}
Multiple importance sampling (MIS) methods use a set of proposal distributions from which samples are drawn.
Each sample is then assigned an importance weight that can be obtained according to different strategies.
This work is motivated by the trade-off between variance reduction and computational complexity of the different approaches (classical vs. deterministic mixture) available for the weight calculation.
A new method that achieves an efficient compromise between both factors is introduced in this paper.
It is based on forming a partition of the set of proposal distributions and computing the weights accordingly.
Computer simulations show the excellent performance of the associated \mbox{\emph{partial deterministic mixture} MIS estimator.}

\end{abstract}

\IEEEpeerreviewmaketitle

\section{Introduction}
\label{sec:intro}

Importance sampling (IS) methods approximate statistical moments of a variable of interest by sets of samples, drawn from a proposal distribution different from the targeted one, and weights, assigned to the samples in order to measure their adequacy in approximating the target \cite{Robert04,Liu04b}.
In its standard form, IS uses one proposal distribution from which all samples are drawn.
However, a more powerful strategy to reduce the variance of the estimators consists of using a set of different proposal distributions.
This is the basis of \emph{multiple importance sampling} (MIS) techniques \cite{Veach95}.
The samples drawn from the different proposals in MIS are then assigned weights proportional to the ratio between the target and the proposal densities evaluated at the sample values.
Several strategies for calculating the weights have been considered, depending on the way in which the proposal evaluation in the denominator is interpreted.
In the standard MIS, the evaluated proposal is exactly the one from which the sample was generated.
This constitutes the simplest method in terms of computational complexity.
A different approach, referred to as \emph{deterministic mixture} (DM) MIS, interprets the set of generating proposals as a whole mixture distribution and calculates the weight of a given sample by considering the entire mixture as the global proposal \cite{Owen00}.
This method attains a variance reduction at the expense of an increase in the computational load \cite{APISvixra,APIS-TSP2015}.

In this work, we propose a novel MIS method that provides an efficient tradeoff in terms of computational complexity and variance reduction of the associated IS estimators.
The approach is based on creating a partition of the set of available proposals and considering that each partitioned set constitutes a mixture distribution.
A sample drawn from a mixand in one of the partitions (mixtures) is then assigned a weight that only accounts for that particular mixture, instead of the entire composite mixture as in the full DM-MIS.
A remarkable reduction in computational complexity is achieved by this approach, while the variance of the associated partial DM-MIS estimator remains comparable to that of the full DM-MIS estimator.
The method can not only applied to IS methods with static distributions (i.e., characterized by fixed parameters), but also to IS methods that adapt the parameters of the proposal distribution in an iterative way (i.e., adaptive IS (AIS) methods \cite{Cappe04,CORNUET12,APIS14}).
Computer simulations show the excellent performance of the proposed scheme in terms of variance reduction for a given computational load.
\section{Problem statement and background}
\label{sec:problem}

In many applications, the interest lies in obtaining the posterior density function (pdf) of set of unknown parameters given the observed data.
Mathematically, denoting the vector of unknowns as  ${\bf x}\in \mathcal{D}\subseteq \mathbb{R}^n$ and the observed data as ${\bf y}\in \mathbb{R}^d$, the pdf is defined as
\begin{equation}
	\normalized{\pi}({\bf x}| {\bf y})
		= \frac{\ell({\bf y}|{\bf x}) g({\bf x})}{Z({\bf y})} \propto \pi({\bf x}|{\bf y})=\ell({\bf y}|{\bf x}) g({\bf x}),
\label{eq_posterior}
\end{equation}
where $\ell({\bf y}|{\bf x})$ is the likelihood function, $g({\bf x})$ is the prior pdf, and $Z({\bf y})$ is the normalization factor.\footnote{From now on, we remove the dependence on ${\bf y}$ to simplify the notation.}
The computation of a particular moment of  ${\bf x}$ is then given by
\begin{equation}
	I = \frac{1}{Z} \int_{\mathcal{D}} f({\bf x}) \pi({\bf x}) d{\bf x},
\label{eq_integral}
\end{equation}
where $f(\cdot)$ can be any integrable function of ${\bf x}$.
In many practical scenarios, we cannot obtain an analytical solution of \eqref{eq_integral} and Monte Carlo methods are used to approximate it.

\subsection{Importance Sampling (IS)}

Let us consider $N$ samples (${\bf x}_1,\ldots, {\bf x}_N$) drawn from a proposal pdf, $q({\bf x})$, with heavier tails than the target, $\pi({\bf x})$.
The samples have associated importance weights given by
\begin{equation} 
	w_i= \frac{\pi({\bf x}_i)}{{q({\bf x}_i)}}, \quad i=1,\ldots,N.
\label{is_weights_static}
\end{equation} 
Using the samples and weights, the moment of interest can be approximated as
\begin{equation}
	\hat{I}_{\textrm{IS}} = \frac{1}{\sum_{j=1}^N w_j} \sum_{i=1}^N w_i  f({\bf x}_i) = \frac{1}{N\hat{Z}}  \sum_{i=1}^N w_i f({\bf x}_i),
\label{eq_partial_estimator_static}
\end{equation}
where $\hat{Z}=\frac{1}{N}\sum_{j=1}^N w_j$ is an unbiased estimator of $Z=\int_{\mathcal{D}} \pi({\bf x}) d{\bf x}$ \cite{Robert04}.
Note that Eq. \eqref{eq_partial_estimator_static} always provides a consistent estimator of $I$, but its variance is directly related to the discrepancy between $\pi({\bf x})|f({\bf x})|$ and $q(\x)$  (for a specific choice of $f$) or to the mismatch between the target $\pi({\bf x})$ and the proposal $q({\bf x})$ (for a general and arbitrary $f$) \cite{Robert04, kahn1953methods}.

\subsection{Multiple Importance Sampling (MIS)}

Finding a good proposal pdf, $q(\x)$, is critical and can also be very challenging \cite{Owen00}.
An alternative and more robust strategy consists of using a population of different proposal pdfs.
This scheme is often known in the literature as {\it multiple importance sampling} (MIS)  \cite{Cappe04, CORNUET12, APIS14}.
Let us consider a set of $N$ (normalized) proposal pdfs, $q_1({\bf x}), \ldots, q_N({\bf x})$, and let us assume that exactly one sample is drawn from each of them, i.e., $\x_i \sim q_i(\x)$, $i=1,...,N$.
The importance weights associated to these samples can then be obtained according to one of the following strategies:
\begin{itemize}
\medskip
\item $\mbox{{\it Standard} MIS:}\;\;\;  w_i= \frac{\pi({\bf x}_i)}{{q_i({\bf x}_i)}}, \quad i=1,\ldots, N. $ \label{weights_SIS}
\medskip
\item {\it Deterministic mixture MIS} (DM-MIS) \cite{Owen00}:
\begin{equation} 
	w_i=\frac{\pi({\bf x}_i)}{\psi({\bf x}_i)}=\frac{\pi({\bf x}_i)}{\frac{1}{N}\sum_{j=1}^{N}q_j({\bf x}_i)}, \quad i=1,\ldots, N,
\label{f_dm_weights_static}
\end{equation}
where $\psi({\bf x})=\frac{1}{N}\sum_{j=1}^{N}q_j({\bf x})$ is the mixture pdf, composed of all the proposal pdfs.
This approach interprets the complete set of samples, $\{{\bf x}_i\}_{i=1}^N$, as being distributed according to the mixture $\psi({\bf x})$, i.e., $\{{\bf x}_1,\ldots, {\bf x}_N\} \sim \psi({\bf x})$.
See Appendix \ref{LucaApp} for further details. 

\end{itemize}

In both  cases, the consistency of the estimators is ensured.
The main advantage of the DM-MIS weights is that they yield more stable and efficient estimators \cite{Owen00}, i.e., with less variance (as proved in Appendix \ref{appendix_varianceDM}).
However, the DM-MIS estimator is computationally more expensive, as it requires $N$ evaluations of the proposal to obtain each weight instead of just one.\footnote{Note that the number of evaluations of the target $\pi({\bf x})$ is the same regardless of whether the weights are calculated according to \eqref{is_weights_static} or \eqref{f_dm_weights_static}.}
In some practical scenarios, this additional load may be excessive (especially for large values of $N$) and alternative efficient solutions must be developed.
 
\subsection{Adaptive Importance Sampling (AIS)}

In order to decrease the mismatch between the proposal and target pdfs, there are several Monte Carlo methods that iteratively adapt the parameters of the proposal pdf using the information of the past samples \cite{Cappe04,CORNUET12,APIS14}.
In this scenario, we have a set of proposal pdfs $\{q_j^{(t)}({\x}), j=1,\cdots,J\}_{t=1}^T$, where the superscript $t$ indicates the iteration index and $T$ is the total number of adaptation steps.
Some of these well-known AIS methods, which are also based on MIS, are Population Monte Carlo (PMC) and its variants \cite{Cappe04,Douc07b,Cappe08,bugallo2009marginalized}, adaptive multiple importance sampling (AMIS) \cite{CORNUET12}, and adaptive population importance sampling (APIS) \cite{APISvixra,APIS14}.

\section{Partial Deterministic Mixture approach}
\label{section_pDM_static}

In this section we develop a \textit{partial} DM-MIS scheme, which groups the proposal pdfs, $\{q_i(\x)\}_{i=1}^N$, into $P$ mixtures composed of $M$ mixands, with $PM=N$ (recall that ${\bf x}_i \sim q_i({\bf x}), i=1,\cdots,N$).\footnote{For the sake of simplicity, we assume that all the mixtures contain the same number of proposal pdfs. However, any strategy that clusters the $N$ proposals into $P$ disjoint mixtures (regardless of their size) is valid.}
Namely, we define a partition of $\{1,\ldots,N\}$ into $P$ disjoint subsets of $M$ indices, $\mathcal{S}_p$ with $p=1,\ldots,P$, s.t.
\begin{equation}
	\mathcal{S}_1 \cup \mathcal{S}_2\cup \ldots \cup \mathcal{S}_P= \{1,\ldots,N\},
\end{equation}
where $\mathcal{S}_k \cap \mathcal{S}_\ell = \emptyset$ for all $k,\ell=1,\ldots,P$ and $k\neq \ell$.
Each subset, $\mathcal{S}_p = \{j_{p,1}, j_{p,2},\ldots, j_{p,M}\}$, contains $M$ indices, $j_{p,m}\in \{1,\ldots,N\}$ for $m=1,\ldots,M$ and $p=1,\ldots, P$.
Following this strategy, the weights of the $p$-th mixture are computed as
\begin{equation}
	w_i = \frac{\pi({\bf x}_{i})}{\psi_p({\bf x}_{i})} 
		= \frac{\pi({\bf x}_{i})}{\frac{1}{M}\sum_{j\in\mathcal{S}_p}q_{{j}}({\bf x}_{i})}, \quad i \in\mathcal{S}_p.
\label{p_dm_weights_static}
\end{equation}

The resulting {\it partial} DM-MIS (p-DM-MIS) estimator is then given by
\begin{equation}
	 \hat{I}_{\textrm{p-DM-MIS}}= \frac{1}{\sum_{j=1}^N w_{j} }  \sum_{i=1}^N w_{i} f({\bf x}_{i}),  
\label{eq_partialDM_estimator}
\end{equation}
which coincides with the expression in \eqref{eq_partial_estimator_static}, but using the weights given by Eq. \eqref{p_dm_weights_static}.
Note that the particular cases $P=1$ and $P=N$ correspond to the {\it full} DM-MIS (f-DM-MIS) and the {\it standard} MIS (s-MIS) approaches, respectively.
%
%
The number of evaluations of the proposal pdfs is then $PM^2$.
Since $N \leq PM^2 = NM \leq N^2$, the computational cost is larger than that of s-MIS approach ($M$ times larger), but lower than that of the f-DM-MIS approach (since $M\leq N$). 

The good performance of the novel approach is ensured by Theorem \ref{variance_theorem} and Corollary \ref{variance_corollary} (see Appendix \ref{appendix_varianceDM}) and can be summarized by the following expression:
\begin{equation}
	\textrm{Var}(\hat{I}_{\textrm{f-DM-MIS}}) \le \textrm{Var}(\hat{I}_{\textrm{p-DM-MIS}}) \le \textrm{Var}(\hat{I}_{\textrm{s-MIS}}),
\label{eq_three_variances_comparison}
\end{equation}
which holds
regardless of the choice of $P$ and the strategy followed to group the original proposals $\{ q_i(\x)\}_{i=1}^N$ into mixtures. 
Therefore, there is a tradeoff between performance and computational cost: using a smaller number of mixtures ($P$) leads to a reduced variance, but at the expense of an increase in the number of evaluations of the proposal densities.

\subsection{Choice of the number of mixtures $P$}

A simple strategy to choose the number of mixtures consists of starting with $P=N$ (which coincides with the standard MIS scheme), computing the corresponding estimator in Eq. \eqref{eq_partialDM_estimator}, and iteratively reducing the number of mixtures $P$ (thus increasing $M=\frac{N}{P}$) while the estimation significantly changes w.r.t. the previous step.
This iterative approach does not require a significant additional computational cost (since the proposal evaluations can be stored and re-used) and results in an efficient and automatic procedure to select $P$.

\subsection{Design of the $P$ mixtures}

Developing an optimal strategy to cluster the proposals into $P$ mixtures is a difficult task, since the number of possible configurations is extremely large.
Indeed, unless this clustering strategy is computationally inexpensive, the additional computational effort might be better invested in decreasing $P$ (thus increasing $M=\frac{N}{P}$ and reducing the variance of the estimators).
Therefore, we propose applying a simple random clustering strategy, where $M=\frac{N}{P}$ different proposals are randomly assigned to each partition.
This approach provides an excellent performance for large values of $M$ (see Table \ref{table_MIS_1}), so there seems to be little room for improvement (except maybe for small values of $M$).
Note that this result is not surprising: randomness is the key element behind compressive sampling \cite{candes2008introduction}, and many randomized clustering algorithms have been developed for applications such as data mining \cite{ng2002clarans}, image processing \cite{moosmann2008randomized} or blind channel identification \cite{luengo2012novel}.

\subsection{Application to AIS schemes}
\label{section_pDM_adaptive}

For the sake of simplicity, we have focused on p-DM-MIS for a static framework, where the parameters of the proposals are fixed.
However, all the previous considerations can be easily applied in AIS schemes, where the proposals are iteratively updated.
In methods like PMC \cite{Cappe04} or APIS \cite{APIS14}, a population of $J$ proposal densities is adapted during $T$ iterations.
At the $t$-th iteration, the $j$-th sample $\x_j^{(t)}$ is drawn from the $j$-th proposal, i.e., $\x_j^{(t)} \sim q_j^{(t)}(\x)$ for $j=1,\ldots,J$ and $t=1,\ldots,T$.
Thus, after $T$ iterations we have $N=JT$ samples drawn from $N=JT$ different proposal pdfs. 

Regardless of the adaptation procedure followed by each algorithm, different strategies can be used to design an efficient DM-MIS estimator when considering all the $N=JT$ proposal pdfs.
Table \ref{Table_weights} summarizes the weight calculation for three well-known AIS methods (PMC, AMIS and APIS), comparing them to the DM-MIS estimators.
We also analyze the complexity in terms of the number of proposal evaluations and the performance in terms of the variance reduction.
We can see that the novel p-DM-MIS approach provides a very good compromise between performance and computational cost.
Finally, it is important to remark that Table \ref{Table_weights} does not take into account the specific adaptive procedures of each algorithm, which can also have a large influence on the final performance.

\begin{table} 
\setlength{\tabcolsep}{2pt}
\def\marginwidth{1.5mm}
\begin{center}
\begin{tabular}{|c@{\hspace{\marginwidth}}|c@{\hspace{\marginwidth}}|c@{\hspace{\marginwidth}}|c@{\hspace{\marginwidth}}|}
\hline
  {\bf Method} &   {\bf Weight calculation} &  {\bf Complexity}  &  {\bf Variance} \\
\hline
\hline
PMC \cite{Cappe04}   &$\frac{\pi({\bf x}_i^{(t)})}{{q_i^{(t)}({\bf x}_i^{(t)})}}$ & Lowest ($JT=N$) & Highest \\
\hline
APIS \cite{APIS14} &$\frac{\pi({\bf x}_i^{(t)})}{\sum_{j=1}^{J}{q_j^{(t)}({\bf x}_i^{(t)})}}$ & Medium ($J^2T=JN$) & Medium \\
 \hline
p-DM-MIS &$\frac{\pi({\bf x}_i^{(t)})}{\sum_{j,\tau \in \mathcal{S}_p}q_{j}^{(\tau)}({\bf x}_i^{(t)})}$ &  Suitable ($PM^2=MN$) & Low \\
\hline
AMIS \cite{CORNUET12} &$\frac{\pi({\bf x}_i^{(t)})}{\sum_{\tau=1}^{T} {q^{(\tau)}({\bf x}_i^{(t)})}}$ & Highest ($T^2=N^2$) & Lowest  \\
\hline
f-DM-MIS &$\frac{\pi({\bf x}_i^{(t)})}{\sum_{\tau=1}^{T} \sum_{j=1}^{J}{q_j^{(\tau)}({\bf x}_i^{(t)})}}$ & Highest ($J^2T^2=N^2$) & Lowest  \\
\hline                                                                                    
\end{tabular}
\end{center}
\caption{Different strategies for weight calculation in AIS algorithms using $J$ proposals per iteration (note that $J=1$ in AMIS), and $T$ iterations.}
\label{Table_weights}
\end{table}

\section{Numerical example}
\label{sec:results}

We consider a bivariate multimodal target pdf, defined as a mixture of $5$ Gaussians:
\begin{equation}
\label{Target1}
\pi({\bf x})=\frac{1}{5}\sum_{i=1}^5 \mathcal{N}({\bf x};{\bf \nu}_i,{\bf \Sigma}_i), \quad {\bf x}\in \mathbb{R}^2,
\end{equation}
where ${\bf \nu}_1=[-10, -10]^{\top}$, ${\bf \nu}_2=[0, 16]^{\top}$, ${\bf \nu}_3=[13, 8]^{\top}$, ${\bf \nu}_4=[-9, 7]^{\top}$, ${\bf \nu}_5=[14, -14]^{\top}$, ${\bf \Sigma}_1=[2, \ 0.6; 0.6, \ 1]$, ${\bf \Sigma}_2=[2, \ -0.4; -0.4, \ 2]$, ${\bf \Sigma}_3=[2, \ 0.8; 0.8, \ 2]$, ${\bf \Sigma}_4=[3, \ 0; 0, \ 0.5]$ and ${\bf \Sigma}_5=[2, \ -0.1; -0.1, \ 2]$.
The goal is to approximate, using some Monte Carlo method, the expected value of ${\bf X}\sim \pi({\bf x})$, i.e., $E[{\bf X}]=\int_{\mathbb{R}^2} {\bf x} \pi({\bf x}) d{\bf x}$ and $Z=\int_{\mathbb{R}^2} \pi({\bf x}) d{\bf x}$.

We apply the MIS algorithm in a setup with $N=4096$ Gaussian proposal pdfs, $\{q_{i}({\bf x})=\mathcal{N}({\bf x};{\bm \mu}_{i},{\bf C}_i)\}_{i=1}^N$, where ${\bm \mu}_i\sim \mathcal{U}([-20,20]\times[-20,20])$ is the randomly chosen location parameter and ${\bf C}_i=\sigma^2{\bf I}_2$, with $\sigma=5$, is the scale matrix.
We proceed as follows.
First, we draw a sample from each proposal.
Then, we compute the corresponding weight according to \eqref{p_dm_weights_static}.
Finally, we build the estimator $\hat{I}_{\textrm{p-DM-MIS}}$ using (\ref{eq_partialDM_estimator}) for different number of mixtures $P=2^k$ for $k=0,1,\ldots,12$ (i.e., $P=1,2,\ldots,2048,4096$).
Since $PM=N$, the number of proposals per mixture is $M = 2^{12-k} = 4096,2048,\ldots,2,1$.
Note that the case $P=1$ (i.e., $M=4096$) corresponds to the f-DM-MIS approach, while $P=4096$ (i.e., $M=1$) corresponds to the s-MIS.
A random assignment of the proposals to the mixtures is performed in all cases.

\begin{figure}
\centering
\includegraphics[width=0.49\textwidth]{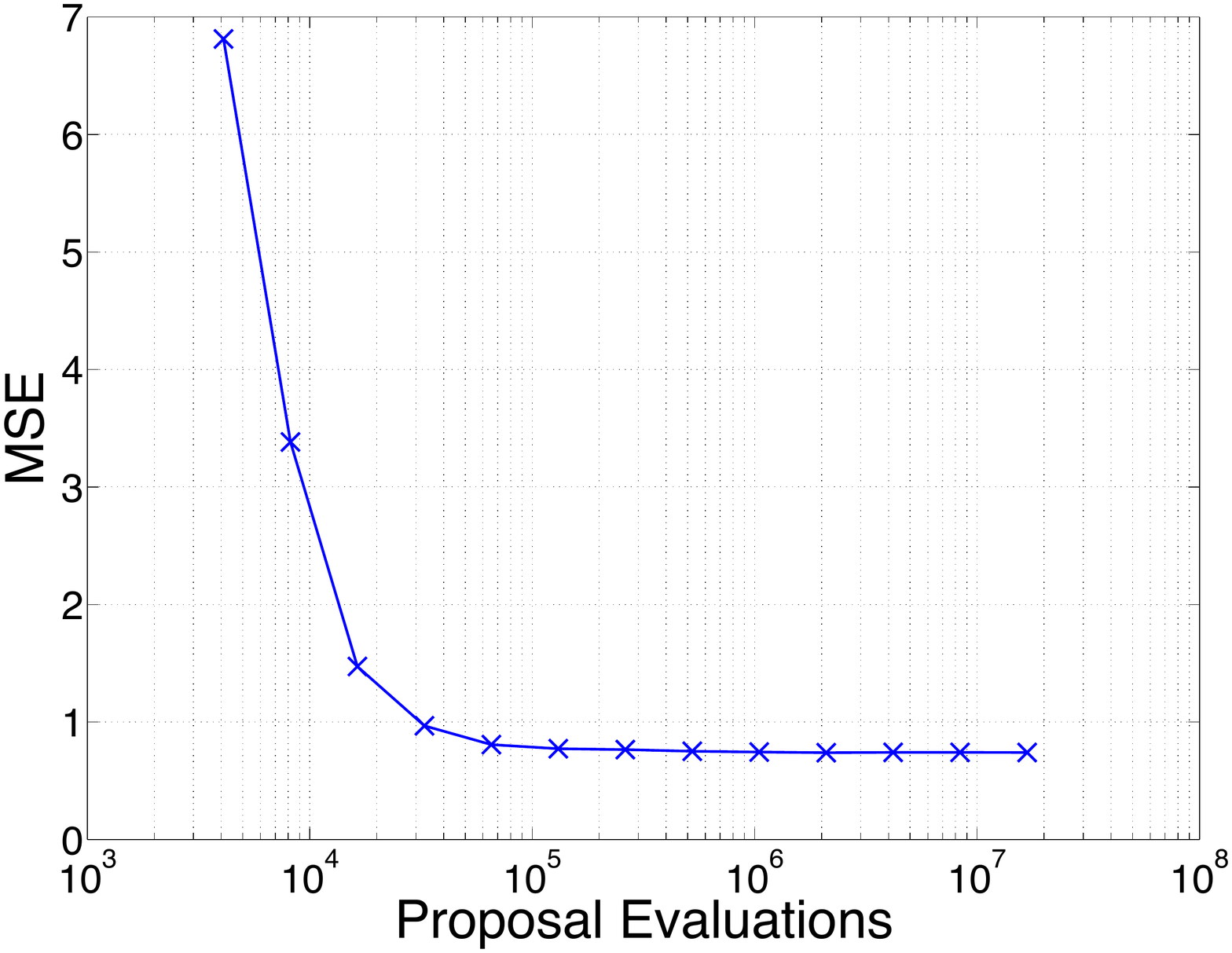}
\caption{MSE in the estimation of the mean of the target w.r.t. the total number of proposal evaluations.}
\label{figure_MIS_1}
\end{figure}

Table \ref{table_MIS_1} shows the mean square error (MSE) in the estimation of the mean of the target $E[{\bf X}]$ (averaged over both dimensions) and the normalizing constant $Z$. 
All results are averaged over $500$ independent runs. 
The last column of Table \ref{table_MIS_1} shows  the total number of proposal evaluations for each value of $M$.
Figure \ref{figure_MIS_1} shows the MSE in the estimation of the mean of the target w.r.t. the total number of proposal evaluations.
The results show that decreasing $P$ reduces the MSE (as expected) at the expense of an increase in the computational cost (measured by the number of proposal evaluations).
However, note that decreasing the number of mixtures below $P=64$ does not improve the performance significantly.
Indeed, the p-DM-MIS with $P=64$ obtains an MSE close to that of the f-DM-MIS estimator while performing $98.4\%$ less proposal evaluations, thus attaining an excellent performance-cost tradeoff.

\begin{table} 
\setlength{\tabcolsep}{2pt}
\def\marginwidth{1.5mm}
\begin{center}
\begin{tabular}{|l@{\hspace{\marginwidth}}|c@{\hspace{\marginwidth}}|c@{\hspace{\marginwidth}}|c@{\hspace{\marginwidth}}|}
\hline
\cline{2-4}
{\bf Parameters} &  MSE($E[{\bf X}]$) &  MSE($Z$) & Evaluations  \\
\hline
\hline
$P=4096$ ($M=1$): s-MIS & 6.8129 & 0.0743 & 4096\\
\hline
$P=2048$ ($M=2$) & 3.3832 & 0.0273 & 8192\\
\hline
$P=1024$ ($M=4$) & 1.4723 & 0.0120 & 16384\\
\hline
$P=512$ ($M=8$) & 0.9678 & 0.0076 & 32768\\
\hline
$P=256$ ($M=16$) & 0.8078 & 0.0064 & 65536\\
\hline
$P=128$ ($M=32$) & 0.7730 & 0.0060 & 131072\\
\hline
$P=64$ ($M=64$) & 0.7648 & 0.0058 & 262144\\
\hline
$P=32$ ($M=128$) & 0.7506 & 0.0058 & 524288\\
\hline
$P=16$ ($M=256$) & 0.7486 & 0.0058 & 1048576\\
\hline
$P=8$ ($M=512$) & 0.7448 & 0.0058 & 2097152\\
\hline
$P=4$ ($M=1024$) & 0.7416 & 0.0058 & 4194304\\
\hline
$P=2$ ($M=2048$) & 0.7414 & 0.0058 & 8388608\\
\hline
$P=1$ ($M=4096$): f-DM-MIS & 0.7406 & 0.0058 & 16777216\\
\hline
\end{tabular}
\end{center}
\caption{MSE in the estimation of the mean and normalizing constant of the target for different values of $P$ and $M$.}
\label{table_MIS_1}
\end{table}

\section{Conclusion}
\label{sec:conclusion}

In this paper we propose a novel approach for the calculation of the weights in multiple importance sampling schemes that provides an efficient tradeoff between computational complexity and variance reduction of the associated estimator.
The proposed scheme is based on constructing a random partition of the set of available proposals and then calculating the weight of each sample locally using only the corresponding subset of the partition.
Computer simulations reveal a very good performance of the method, which is able to attain an excellent performance-computational cost tradeoff.
%

%

\appendices

\section{Drawing samples from a mixture of pdfs}
\label{LucaApp}

Let us consider a mixture of $N$ normalized pdfs with equal weights, i.e.,
\begin{equation}
	\psi({\bf x}) = \frac{1}{N}\sum_{i=1}^{N}{q_i({\bf x})}.
\label{eq:mixture}
\end{equation}
The classical procedure for drawing $N$ samples from $\psi({\bf x})$ is (starting with $k=1$):
\begin{enumerate}
	\item Draw $j^* \in \{1,\ \ldots,\ N\}$ with equal weights $\frac{1}{N}$.
	\item Draw ${\bf x}_{k} \sim q_{j^*}({\bf x})$.
	\item Set $k=k+1$ and repeat until $k=N$.
\end{enumerate}
In this way, each sample ${\bf x}_{k}$ is distributed according to $\psi({\bf x})$ and, as a consequence, $\{{\bf x}_{1},\ \ldots,\ {\bf x}_{N} \} \sim \psi({\bf x})$.
An alternative procedure, more {\it deterministic} than the previous one, consists of the following steps (starting with $i=1$):
\begin{enumerate}
	\item Draw one sample from each $q_{i}({\bf x})$, i.e., ${\bf x}_{i} \sim q_{i}({\bf x})$.
	\item Set $i=i+1$ and repeat until $i = N$.
\end{enumerate}
In this case, we have ${\bf x}_{i} \sim q_{i}({\bf x})$ for $i=1,\dots,N$, {\it but} the joint set is again distributed as $\psi({\bf x})$, i.e.,	$\{{\bf x}_{1}, {\bf x}_{2},\ \ldots,\ {\bf x}_{N}\} \sim \psi({\bf x})$.
This is the underlying theoretical argument of the deterministic mixture (DM) weights approach used throughout this work.
Furthermore, given $M$ indices $j_r\in \{1,\ldots, N\}$ with $r=1,\ldots, M$:
\begin{equation*}
	\{{\bf x}_{j_1}, \ldots, {\bf x}_{j_M}\} \sim \frac{1}{M} q_{j_1}({\bf x})+\ldots+  \frac{1}{M} q_{j_M}({\bf x}).
\end{equation*}

\section{Variance reduction of deterministic mixture weights}
\label{appendix_varianceDM}

\begin{theorem}
\label{variance_theorem}
Consider a normalized target pdf, $\tilde{\pi}({\bf x}) = \frac{1}{Z} \pi({\bf x})$, and $N$ samples drawn from a set of $N$ normalized proposal pdfs (one from each pdf), ${\bf x}_i \sim q_i({\bf x})$ for $i = 1,2,\ldots,N$.
The standard multiple importance sampling (s-MIS) and the full deterministic mixture IS (f-DM-IS) estimators can be expressed as 
\begin{subequations}
\begin{align}
	\hat{I}_{\textrm{s-MIS}} & = \frac{1}{N} \sum_{i=1}^N  \frac{f({\bf x}_i) \pi({\bf x}_i)}{q_i({\bf x}_i)},\label{eq:estMIS}\\
	\hat{I}_{\textrm{f-DM-MIS}} & = \frac{1}{N} \sum_{i=1}^N  \frac{f({\bf x}_i) \pi({\bf x}_i)}{\frac{1}{N}\sum_{j=1}^{N}{q_j({\bf x}_i)}}.\label{eq:estfullDM}
\end{align}
\end{subequations}
The variance of the f-DM-IS estimator is always lower or equal than the variance of the s-MIS estimator, 
\begin{equation}
	\textrm{Var}(\hat{I}_{\textrm{f-DM-IS}}) \le \textrm{Var}(\hat{I}_{\textrm{s-MIS}}).
\label{eq:varIneq}
\end{equation}
\end{theorem}

\noindent
{\bf Proof:} See Appendix A of \cite{APISvixra,APIS-TSP2015}. $\hfill\Box$

\begin{corollary}
\label{variance_corollary}
For the standard MIS, full DM-MIS and partial DM-MIS estimators the following inequality holds:
\begin{equation*}
	\textrm{Var}(\hat{I}_{\textrm{f-DM-MIS}}) \le \textrm{Var}(\hat{I}_{\textrm{p-DM-MIS}}) \le \textrm{Var}(\hat{I}_{\textrm{s-MIS}}).
\end{equation*}
\end{corollary}

\noindent
{\bf Proof:} For the first inequality, note that the full DM-MIS estimator of (\ref{eq:estfullDM}) can also be expressed as
\begin{equation*}
	\hat{I}_{\textrm{f-DM-MIS}} = \frac{1}{N}\sum_{p=1}^{P}{\sum_{i\in\mathcal{S}_p}{\frac{f(\x_i)\pi(\x_i)}{\frac{1}{P}{\sum_{p=1}^{P}{\psi_p(\x_i)}}}}},
\end{equation*}
where 
\begin{equation*}
	\psi_p(\x_i) = \frac{1}{M} \sum_{j\in\mathcal{S}_p}{q_j(\x_i)}
\end{equation*}
is the proposal associated to the $p$-th mixture.
Similarly, the partial DM-MIS estimator is given by
\begin{equation*}
	\hat{I}_{\textrm{p-DM-MIS}} = \frac{1}{N}\sum_{p=1}^{P}{\sum_{i\in\mathcal{S}_p}{\frac{f(\x_i) \pi(\x_i)}{\psi_p(\x_i)}}}.
\end{equation*}
Hence, applying Theorem \ref{variance_theorem} with $\psi_p(\x_i)$ instead of $q_i(\x_i)$ and $\frac{1}{P} \sum_{p=1}^{P}{\psi_p(\x_i)}$ instead of $\frac{1}{N}\sum_{j=1}^{N}{q_j({\bf x}_i)}$, we have $\textrm{Var}(\hat{I}_{\textrm{f-DM-MIS}}) \le \textrm{Var}(\hat{I}_{\textrm{p-DM-MIS}})$.

For the second inequality, following the same approach, the standard MIS estimator of (\ref{eq:estMIS}) can be rewritten as
\begin{equation*}
	\hat{I}_{\textrm{s-MIS}} = \frac{1}{N}\sum_{p=1}^{P}{\sum_{i\in\mathcal{S}_p}{\frac{f(\x_i) \pi(\x_i)}{q_i(\x_i)}}}.
\end{equation*}
Applying Theorem \ref{variance_theorem} with $\psi_p(\x_i) = \frac{1}{P} \sum_{p=1}^{P}{\psi_p(\x_i)}$ instead of $\frac{1}{N}\sum_{j=1}^{N}{q_j({\bf x}_i)}$, we have $\textrm{Var}(\hat{I}_{\textrm{p-DM-MIS}}) \le \textrm{Var}(\hat{I}_{\textrm{s-MIS}})$. $\hfill\Box$

\bibliographystyle{IEEEbib}

\end{document}